# SEGMENTATION OF DYNAMIC CONTRAST-ENHANCED MAGNETIC RESONANCE IMAGES OF THE PROSTATE

**Wuilian Torres***
*wuiliantor@gmail.com*
Centro de Procesamiento Digital de Imágenes, Instituto de Ingeniería
Urb. Monte Elena II, Sartenejas, Baruta, Caracas-Venezuela
**Leonardo Cordero***
*leojcq@gmail.com*
Hospital Universitario, Universidad Central de Venezuela
Ciudad Universitaria, Av. Los Estadios, Urb. Los Chaguaramos, Caracas-Venezuela
**Miguel Martin-Landrove***
*mglmrtn@gmail.com*
Centro de Física Molecular y Médica, Facultad de Ciencias, Universidad Central de Venezuela
Ciudad Universitaria, Av. Los Estadios, Urb. Los Chaguaramos, Caracas-Venezuela
Centro de Visualización Médica, Instituto Nacional de Bioingeniería, Universidad Central de Venezuela, Av. Miguel Otero Silva, Urb. Sebucán, Caracas, Venezuela
**Antonio Rueda-Toicen***
antonio.rueda.toicen@gmail.com
Centro de Visualización Médica, Instituto Nacional de Bioingeniería, Universidad Central de Venezuela, Av. Miguel Otero Silva, Urb. Sebucán, Caracas, Venezuela
Algorithmic Nature Group, LABORES for the Natural and Digital Sciences, Paris, France

*Member of the Physics & Mathematics in Biomedicine Consortium

**Abstract.** *Dynamic Contrast-Enhanced Magnetic Resonance Imaging (DCE-MRI) is a valuable tool to localize, characterize, and evaluate anomalous prostate tissue. The DCE-MRI technique produces ultrafast gradient-echo acquisitions of MRI volumes with high temporal resolution that are generated at regular intervals while the patient receives, in a controlled manner, a paramagnetic contrast agent. Angiogenesis in malignant tumors of the prostate is characterized by aggressive vessel growth that contributes to the rapid evacuation ("wash-out") of the contrast agent. This feature becomes evident in a time series where each voxel exhibits a particular behavior of contrast uptake ("wash-in") and posterior wash-out; this behavior depends on both the properties of the tissue represented by the voxel and its degree of vascularization. In this work, we propose a segmentation method that groups together neighboring voxels with similar contrast*

*wash-out responses. The segmentation algorithm uses a region growing technique that is a variant of the "GrowCut" cellular automaton. This cellular automaton has cells associated with each voxel in the volume, and every cell in the automaton evolves iteratively according to its relationship with its neighboring cells. Initially, the cellular automaton is given "seed" cells, which are determined in an automatic manner through morphological filters that identify homogeneous regions in the volume. Seed cells are representatives of the clinically relevant types of tissues in the prostate. Each cell in the automaton has three parameters: a label that identifies the type of tissue in the associated voxel, a vector with the values of the DCE-MRI time series, and a coefficient called "strength" with values between 0 and 1 that is interpreted as the probability with which the cell belongs to its assigned label. Every non-seed cell can modify its state; this occurs when each cell is attacked by its neighbors with a strength of attack that is inversely proportional to the similarity of the values of the time series between the cells. If the strength of the attacked cell is less than the strength of the attack of one of its neighbors, the state of the attacked cell changes and it takes the label of the attacking cell. The attacked cell also updates its strength making it equal to the strength of the attack with which it was conquered. To perform a clinical validation of the resulting segmentations, we used various cases from the database of The Cancer Imaging Archive (TCIA), National Cancer Institute (NCI) [5, 6].*

## 1. INTRODUCTION

Segmentation groups together regions of an image according to functional or structural criteria. In this work we use functional criteria to segment and discriminate prostate tissue according to its degree of vascularity. For this we use Dynamic Contrast-Enhanced Magnetic Resonance Imaging (DCE-MRI). Magnetic resonance imaging (MRI) is an essential tool to diagnose and classify tumors due to the information that it provides related to their morphology and interaction with their surrounding tissue. The DCE-MRI technique was pioneered in 1991 and is being commonly used clinically both for diagnostic purposes and while supervising the evolution of tumors during treatment.

Prostate cancer is one of the most prevalent type of cancers, and its early detection is important to permit treatment and survival. MRI T2-weighted images (T2WI) are commonly used for the evaluation of the prostate, and in these images it's possible to distinguish 2 regions: the central gland and its periphery. The central gland has signals of mid intensity and heterogeneous texture, while the periphery produces signals of high intensity. Eighty percent of prostatic carcinomas develop on the periphery of the prostate. Tissues affected by prostate cancer have abnormally low signal intensity levels on T2WI [1].

Usually, the analysis of prostate magnetic resonance images requires the inspection and comparison of multiple sequences. The identification of tissue is performed manually by expert radiologists who base their judgements on their knowledge of the anatomy of the tissue. The manual identification of the edges of the prostate in an MRI series can be difficult, slow, and error-prone. There is also inter-variability in the segmentations produced by different experts.

The development of algorithms to perform automatic segmentation is challenging mainly due to the variability in the size and shape of the prostate between patients and the variability of the intensity of the signal in tissues [2].

In this work, we propose an emergent computing technique for the automatic segmentation of prostate tissue that considers both topological and functional criteria in DCE-MRI.

## 2. DYNAMIC CONTRAST-ENHANCED MAGNETIC RESONANCE IMAGING

Images acquired through magnetic resonance contain information about the morphology of a tumor and the relationship of the malignancy with surrounding tissue [1]. The growth of new blood vessels (angiogenesis) is essential for the growth and development of tumors, and an indicator of the degree and progression of many tumors is the density of new vascularization. The vascularization of a tumor can be estimated observing its blood flow, and this behavior can be observed through the administration of a paramagnetic contrast agent; variations of perfusion in tissues of the contrast agent are proportional to their vascularization. Dynamic contrast-enhanced magnetic resonance images are acquired as an ultrafast echo gradient sequence before, during, and after administering the contrast agent to the bloodstream. The contrast is injected at a rate of 2-4 mL/s, followed by 20 mL of saline solution. Ideally 3D volumes are acquired every 5 seconds during a period of 5 to 10 minutes to detect the maximum peak of uptake. It's desirable to acquire the highest possible number of dynamic series to obtain perfusion curves with high detail.

DCE-MRI allows the direct observation of the dynamics of the contrast agent and its neighboring tissues. Parametric methods such as the kinetic models of Tofts [3, 4] are commonly used in clinical studies to estimate the transference of plasma in the extravascular-extracellular space. Non-parametric methods aim to identify tissues through the analysis of the dynamics of the contrast agent; the temporal evolution of every voxel of the MRI can be plotted on a Signal-Intensity $SI(t)$ curve (Eq. 1), where $S(t)$ is the value of the voxel's signal at time $t$, and $S(t_0)$ the intensity value of the voxel before injecting the contrast agent.

$$SI(t) = \frac{S(t) - S(t_0)}{S(t_0)} \qquad (1)$$

Figure 1 shows the typical Signal-Intensity curve; the first value is obtained with the first MRI sequence before injecting the contrast at $t_0$, the interval while the contrast is injected to the bloodstream until it reaches its maximum concentration at $t_p$ is called "wash-in", afterwards its presence must decrease slowly during the interval called "wash-out". $S(t)$ allows for the estimation of indices related to the vascularization of the tissues. On tumoral tissues, the amount of blood vessels is much larger, being much more permeable than normal tissues. Blood vessels in tumors are also larger and more disorganized than blood vessels in normal anatomy. On Figure 1b we present the behavior of three patterns during the wash-out phase. The type 1 pattern, shown in blue, shows the progressive increase of the contrast agent that's typical in healthy tissue, while the green and red patterns correspond to an uncertain and pathological condition, respectively.

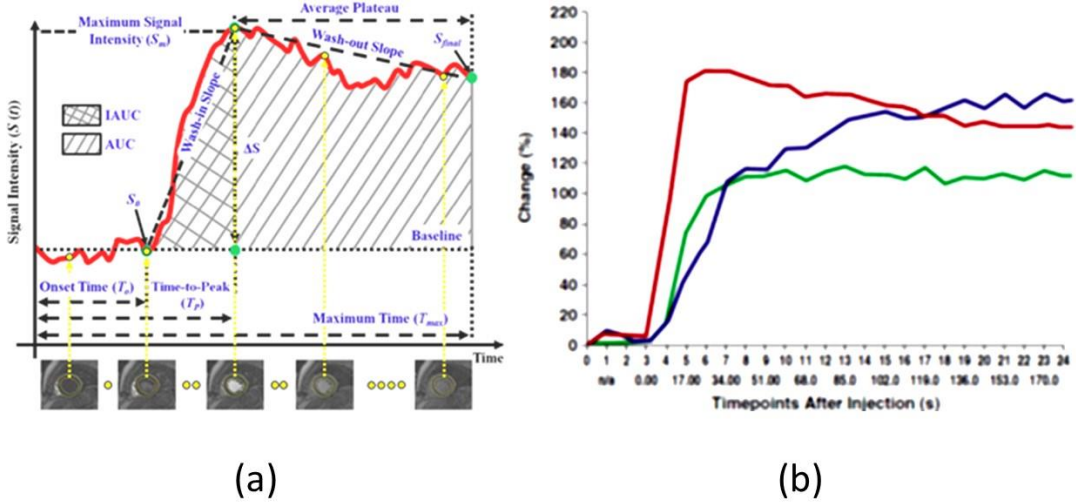

Figure 1- (a) Typical signal-intensity curve. (b) Characteristic patterns of various types of tissue

## 3. DCE-MRI DATASETS

To develop this work, we used public DCE-MRI datasets from The Cancer Imaging Archive (TCIA) [5, 6]. T1-weighted and T2-weighted MRI of cases of prostate cancers were acquired using a Phillips Achieva resonator of 1.5T using an endorectal coil. Patients were administered a paramagnetic contrast agent of gadolinium-DTPA.

## 4. SEGMENTATION WITH THE GROWCUT CELLULAR AUTOMATON

Unsupervised segmentation through the GrowCut cellular automaton has been used to analyze multispectral satellite images [7] and multimodal MRI [8]. In this work, we propose an adaptation of the techniques designed for segmenting multispectral and multimodal images on the segmentation of multitemporal DCE-MRI.

The GrowCut cellular automaton was originally proposed by Vezhnevets and Konouchine [9]. This cellular automaton is represented mathematically by the tuple $(Z^n, S, N, \delta)$, where $Z^n$ is an n-dimensional lattice of regular cells, S is the set of possible states, $N$ is the neighborhood of the cells, and $\delta$ is the transition function used to determine the state of the cell at the evolution step $t + 1$ according to the state of the cells in the neighborhood at the evolution step t. The two most commonly used neighborhood systems are the Von Neumann ("4-connected pixels" in 2D) and the Moore neighborhoods ("8-connected pixels" in 2D). The state at evolution step t of a given cell $p$ is a tuple $(l_p, \theta_p, \vec{C}_p)$, where $l_p$ is the label of the cell, $\theta_p$ is a value in the interval $[0, 1]$ that represents the strength of the cell, and $\vec{C}p$ is the vector with the characteristic attributes of the cell. The evolution rule of the cellular automaton uses the monotonous decreasing function $g$ to determine the strength with which the cell $p$ is attacked by the neighboring cell $q$, Eq. (2) defines the attack strength used to evolve the state of cell while it's being attacked by the cell $q$. The attack strength is proportional to the closeness between the characteristic attribute vectors of the cells.

$$g(|\vec{C_p} - \vec{C_q}|)\, \theta_q > \theta_p \quad with: g(x) = 1 - \frac{x}{\max(\vec{C})} \qquad (2)$$

When one or more of the neighbors of cell $p$ satisfies the condition defined in Eq. (1), the state of $p$ is updated according to the label and strength of the strongest attack by the conquering cell $q$.

In this work, we consider the vector $\vec{C_p}$ as constituted by the time series of the signal-intensity in the DCE-MRI in the wash-out phase, starting from the peak uptake step $T_p$.

## 5. IMAGE SEGMENTATION PIPELINE

Each one of the voxels in the DCE-MRI volume is associated with a cell in the automaton. At its initial state ($t = 0$), every non-seed cell in the automaton has strength zero and a null label. It's necessary to identify in the DCE-MRI the voxels that represent the different type of tissues that are present in the studied volume, and that are characterized by their behavior during the wash-out phase. These voxels are seed cells that define the initial state of the automaton and receive a characteristic label alongside the maximum possible strength of 1.

The procedure used to determine the seed cells is the following:

1. Determine the Signal-Intensity vectors for the voxels.
2. Determine a mask that defines the location of the prostate.
3. Identify the instant when the cell takes the maximum value $T_p$ and construct the vector of attributes $\vec{C}$.
4. Calculate the morphological gradient to find homogenous zones according to $\vec{C}$.
5. Simplify the data corresponding to the homogenous zones, grouping these in N types of possible tissues through k-means clustering.
6. Assign to every cell in the homogenous zone the label of the corresponding group. These cells are also assigned a strength equal to 1.

### 5.1 Signal-Intensity Information

Figure 2 shows the images corresponding to one of the volume slices in the DCE-MRI sequence. The first image shows a slice of the magnetic resonance image before injecting the contrast. The images are acquired at regular intervals while the contrast agent is injected (wash-in) and when it's evacuated (wash-out).

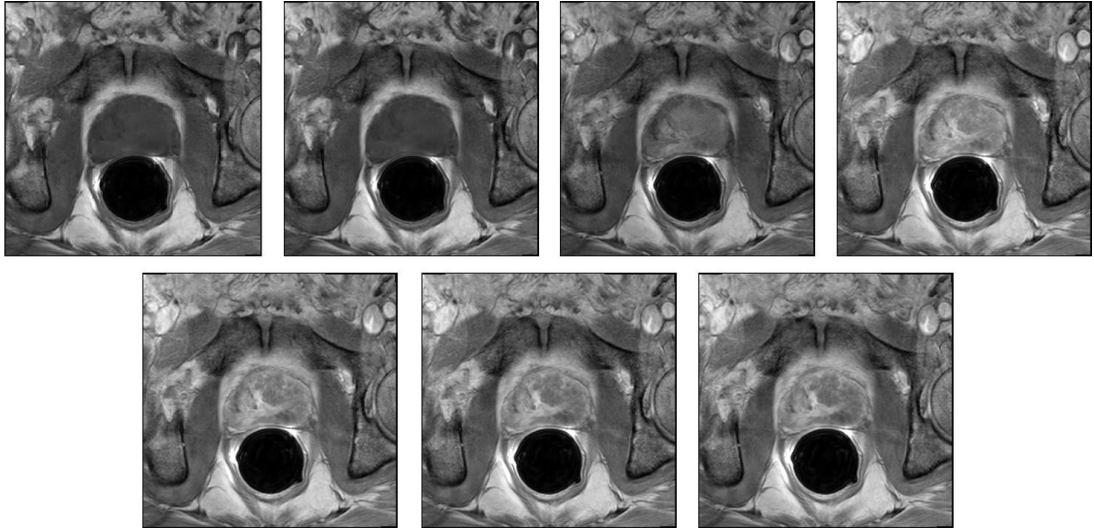

Figure 2 - DCE-MRI Sequence, the upper left corner is a slice of MRI before injecting the contrast, the following images are acquired after injecting the contrast agent.

Signal-intensity information is determined on every voxel using Eq.1; the resulting images can be observed in Figure 3.

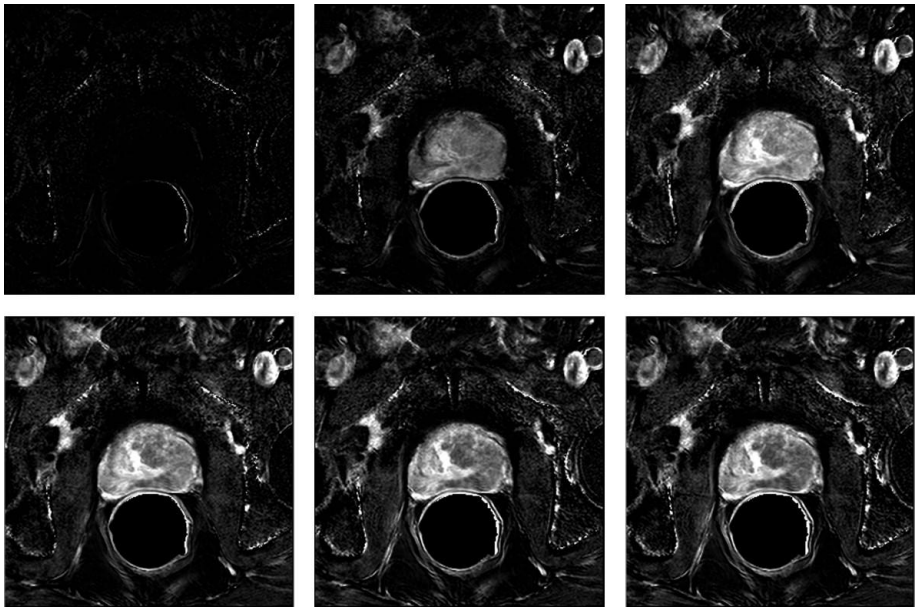

Figure 3 – Signal-Intensity Sequences

**5.2 Region of Interest Masking**

We define a region of interest (ROI) mask to guide the analysis. The DCE-MR images from the TCIA dataset were obtained using an endorectal coil; consequently, in these images the prostate is located just on top of the coil. The ROI mask for the prostate is obtained in two steps. First we identify the position of the coil using the watershed morphological operator. In the second step, we use the gradient of one of the signal-intensity images, imposing the obtained minima as internal markers for the prostate based on their position with respect to the coil; we apply again the watershed operator and obtain the contour of the prostate in the whole volume. Fig 4a shows the mask in one of the slices and Figure 4b shows the gradient within the ROI.

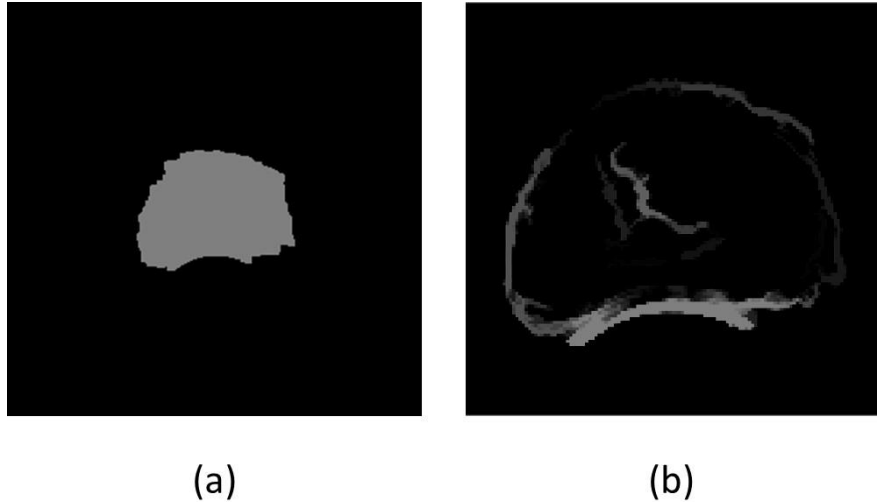

(a)  (b)

Figure 4 – (a) Mask of the prostate. (b) Gradient region within the mask

**5.3 Determining the start of the wash-out phase and attributes for the automaton's cells**

As shown in Figure1 the average slope of signal-intensity curve is higher during the injection of the contrast agent. We evaluate the average variation of the slope and define $T_p$ as the time step where an abrupt change occurs in the behavior of the curve. The vector of attributes for each of the automaton's cells corresponds to the values of signal-enhancement taken from $T_p$, representing the start of the wash-out phase.

**5.4 Seed Location and Labeling**

Each type of tissue in the prostate is represented by voxels with a functional behavior as the one described on Figure 1. Signal intensity is more homogeneous at the center of the tissue and more heterogeneous at its periphery. Consequently, the gradient is lower at the central voxels of the tissues. We apply a morphological operator of regional minima to identify the voxels with a relatively low gradient. These zones become the cellular automaton's seeds.

The assignment of labels is done clustering cells according to their attribute similarity. The Density-Based Spatial Clustering of Applications with Noise (DBSCAN) algorithm is used to determine the N groups of data that have higher attribute similarity. To reduce compute time, we apply DBSCAN on the first 2 principal components of the data associated to the seed cells. The parameters of the DBSCAN are modified until we obtain a number of clusters that represents the amount of relevant tissues in the ROI, in this case we consider 10 different types of tissues. Figure 5 shows the location of the seeds with coloring according to the label.

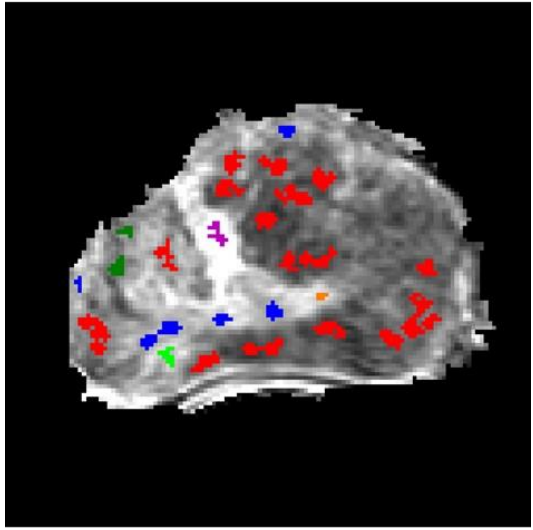

Figure 5 – Seeds at the cellular automaton's initial state ($t = 0$)

## 5.5 Cellular Automaton Evolution and Segmentation

The cellular automaton's seed cells have as label the group at which they were assigned and have the maximum possible strength: 1.These cells won't be able to change state and, following the cellular automaton's evolution rule, will attack neighboring cells, attempting to conquer them and make them take their label. The cellular automaton will continue evolving as long as cells keep updating their state. The evolution is finished when zero updates occur in the lattice. Figure 6a shows the ROI in one of the DCE-MRI slices; Figure 6b shows the segmentation that's achieved using the seed cells described in the previous paragraph. Figure 6c shows contours of the segmentation superimposed on the original image.

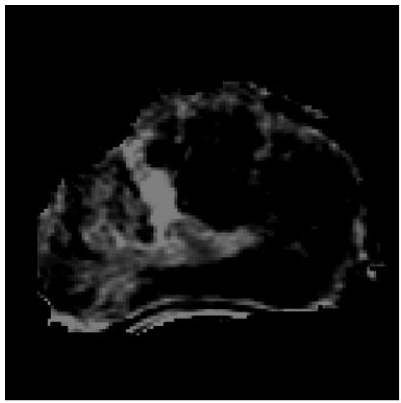
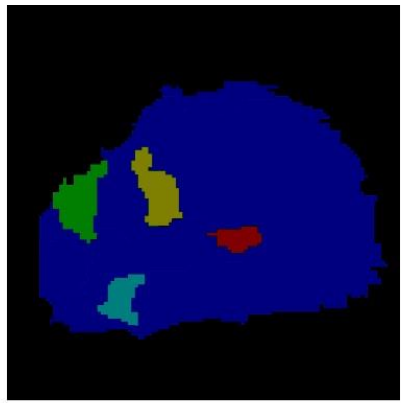
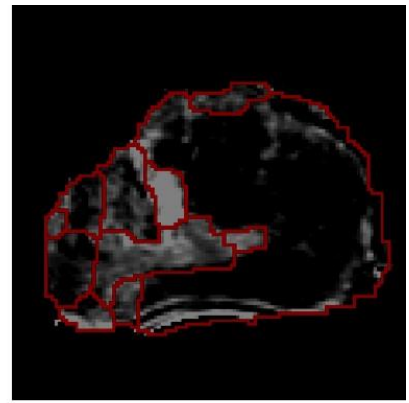

(a) (b) (c)

Figure 6 - (a) Prostate ROI. (b) Segmentation by the cellular automaton. (c) Borders of segments superimposed on the original prostate ROI

## 6. SEGMENTATION RESULTS

On Figure 7, we present segmentation results in a series of images of the TCIA [5, 6]. The left column shows one of the slices of the DCE-MRI volume, the center column shows the location and labeling of the segmentation seeds in the slice, and the right column shows the original image with superimposed segmentation borders. The borders delimit regions according to their similarity in signal-intensity values.

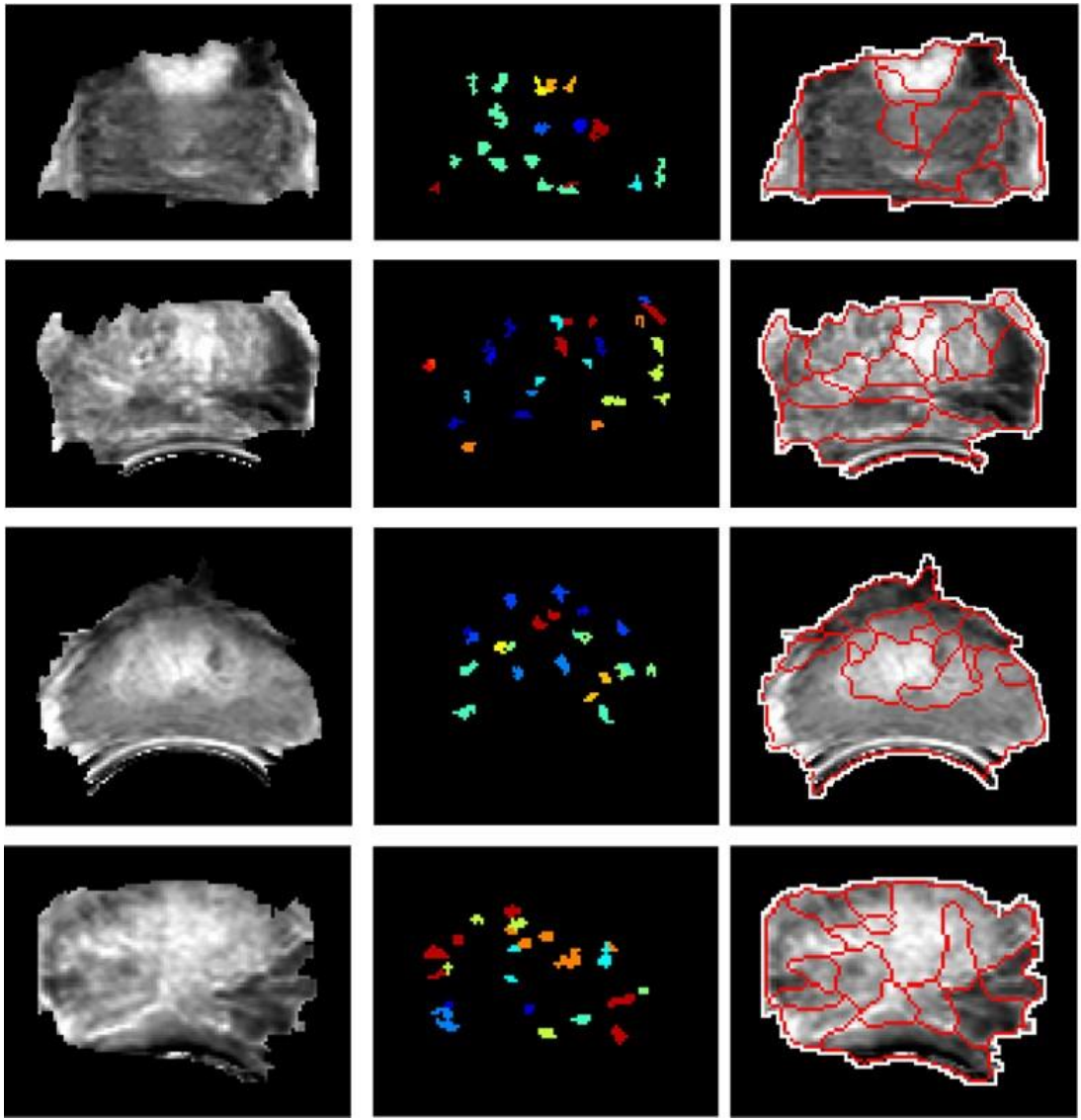

Figure 7 - Segmentation of DCE-RMI volumes from the TCIA database [5, 6]

## 7. CONCLUSIONS AND FURTHER WORK

Even though the seed selection and segmentation process shows promising results, it's necessary to improve the pipeline. In particular, it's important to preprocess the image to delimit the prostate with a higher precision and to include as a feature for the algorithm the correlation between the different volume slices. We also wish to include more functional criteria in the seed selection process that guides the cellular automaton's evolution.

*Acknowledgements*

The authors would like to thank the research support given by the Physics and Mathematics in Biomedicine Consortium during the realization of this work.